\documentclass[12pt, onecolumn, draftclsnofoot,peerreview]{IEEEtran}

\usepackage{cite}
\usepackage{pslatex}
\usepackage{epsfig}
\usepackage{amsmath}
\usepackage{amssymb}

\usepackage{array}
\usepackage{threeparttable}
\usepackage{url}
\usepackage{dsfont}

\newtheorem{theorem}{Theorem}
\newtheorem{proposition}{Proposition}

\newtheorem{definition}{Definition}
\newtheorem{corollary}{Corollary}

\def \tb#1#2{\mathop{#1 \vphantom{\sum}}\limits_{\displaystyle #2}}
\def \tbsmall #1#2{\mathop{#1 \vphantom{\sum}}\limits_{\scriptstyle #2}}

\DeclareMathAlphabet{\mathpzc}{OT1}{pzc}{m}{it}

\IEEEoverridecommandlockouts 
\begin{document}

\title{Cognitive Radio with Partial Channel State Information at the Transmitter}

\author{
\authorblockN{Pin-Hsun Lin, Shih-Chun Lin, Chung-Pi Lee, and Hsuan-Jung
Su}
\authorblockA{}
\thanks{Pin-Hsun Lin, Chung-Pi Lee, and Hsuan-Jung
Su are with the Department of Electrical Engineering and Graduate
Institute of Communication Engineering, National Taiwan
University, Taipei, Taiwan. Shih-Chun Lin is with the Institute of
Communications Engineering, National Tsing Hua University, Taipei,
Taiwan. Email: \{f89921145, r94942026\}@ntu.edu.tw,
hjsu@cc.ntu.edu.tw, and sclin2@ntu.edu.tw. This work was presented
in part at the IEEE International Conference on Communications,
May 2008, and at the Annual Conference on Information Sciences and
Systems, Mar. 2008 and 2009.}
 }

\maketitle 

\vspace{-3mm} \maketitle \IEEEpeerreviewmaketitle
\begin{abstract}
In this paper, we present the cognitive radio system design with
partial channel state information known at the transmitter (CSIT).
We replace the dirty paper coding (DPC) used in the cognitive
radio with full CSIT by the linear assignment Gel'fand-Pinsker
coding (LA-GPC), which can utilize the limited knowledge of the
channel more efficiently. Based on the achievable rate derived
from the LA-GPC, two optimization problems under the fast and slow
fading channels are formulated. We derive semi-analytical
solutions to find the relaying ratios and precoding coefficients.
The critical observation is that the complex rate functions in
these problems are closely related to ratios of quadratic form.
Simulation results show that the proposed semi-analytical
solutions perform close to the optimal solutions found by
brute-force search, and outperform the systems based on naive DPC.
Asymptotic analysis also shows that these solutions converge to
the optimal ones solved with full CSIT when the $K$-factor of
Rician channel approaches infinity. Moreover, a new coding scheme
is proposed to implement the LA-GPC in practice. Simulation
results show that the proposed practical coding scheme can
efficiently reach the theoretical rate performance.
\end{abstract}

\vspace{-3mm}
\section{Introduction}
As the demand for high data rate steadily increases, efficient
spectrum usage becomes a critical issue. Recent measurements from
the Federal Communications Commission (FCC) have indicated that 90
percent of the time, many licensed frequency bands remain unused.
A radio technology that promises to solve such problems is the
cognitive radio (CR) \cite{Mitola_CR}. This technology is capable
of dynamically sensing and locating unused spectrum segments in a
target spectrum pool, and communicating via the unused spectrum
segments without causing harmful interference to the primary
users. The primary users are defined as those of existing
commercial standards. If a primary user demands the channel, the
CR user should vacate and find an alternative one. Assuming that
there is perfect channel state information at the transmitter
(CSIT), Devroye \textit{et. al.} in
\cite{Devroye_CR_achievable_rate} proposed the concept of
\textit{interference mitigation} CR in which the CR and primary
users can simultaneously transmit at the same time and frequency
bands. The CR transmitter not only transmits its own signal but
also relays the primary users'. This is the key to keeping the
primary users' transmission rates unchanged under the coexistence
of CR users since the CR users' signals may interfere with the
primary users'. For the CR receiver, signals from the primary
transmitters and those relayed by the CR transmitter are
interferences. These interferences are non-causally known at the
CR transmitter and can be precoded with dirty paper coding (DPC)
\cite{CostaDPC} such that the CR receiver can decode as if these
interferences do not exist. The validity of the  non-causally
known interferences has been discussed in \cite{Viswanath_CR}. DPC
is a promising precoding technique for cancelling arbitrary
interferences perfectly known only at the transmitter but not at
the receiver. Devroye \textit{et. al.}
\cite{Devroye_CR_achievable_rate} then derived the achievable rate
of the DPC based interference mitigation CR system. In
\cite{Viswanath_CR}, Jovicic and Viswanath further proved that the
DPC based CR system indeed achieves the capacity.

Due to limited feedback bandwidth and the delay caused by channel
estimation, it is more practical to consider the case with only
partial CSIT. Unfortunately, the interference-free rate achieved by DPC
heavily relies on perfectly known CSIT. It has
been observed that naively applying DPC in fading channels with partial CSIT may
cause a significant performance loss
\cite{DPC_BennatanfadingPaper}. In this paper, we focus on the CR
system design for the cases where only the fading channel
statistics are available at both the primary and CR transmitters.
The main contributions of this paper are as following. We use a coding scheme
more general than the DPC, termed the linear-assignment
\textit{Gel'fand-Pinsker coding}
 (LA-GPC) \cite{Gelfand_Pinsker} to derive the achievable rate of the CR
channel. Based on the achievable rate, we then propose design
methods to optimize the performance of the CR system with only the
statistics of CSIT for both fast and slow fading scenarios
\cite{Book_Tse}. The goal is to keep the primary user's rate
unchanged while maximizing the CR user's rate. The considered
problems are non-convex and are almost analytically intractable.
Therefore, we provide semi-analytical solutions to the optimal
relaying ratio of the CR transmitter in fast and slow fading
channels, respectively. These solutions are obtained by expressing
the primary user's rate constraint in the \textit{ratio of
quadratic form}, then applying the techniques of moments of
quadratic forms and the \textit{deterministic approximations of
probability inequalities} \cite{Pinter}. To optimize the CR user's
own rate, we also provide semi-analytical solutions to the
precoding coefficients in fast and slow fading channels,
respectively. In slow fading channels, we show that the complex
objective function can be recast as a compound function of the
incomplete gamma function from which the precoding coefficient can
be easily numerically solved. Asymptotic analysis is also given to
show that these solutions converge to the optimal ones solved with
full CSIT when the $K$-factor of Rician channel approaches
infinity. This result verifies the correctness of the proposed
methods. Simulation results show that the proposed semi-analytical
methods perform well compared to the optimal solutions found by
brute-force search under various channel conditions. In addition
to the theoretical results based on unstructured random Gaussian
codebooks, we also use nested-lattice codebooks with lattice
decoding to implement the precoding in practice \cite{Pslin_CR2}.
Simulation results show that the nested-lattice coding scheme can
efficiently approach the theoretical achievable rate.

This paper is organized as follows. Section \ref{Sec_system_model}
gives the system model of the CR system with partial CSIT.
Review of the LA-GPC and its application to the partial CSIT CR
system are also given in this section. The proposed parameter
design for fast and slow fading channels are discussed in
Section \ref{Sec_fast_fading} and Section \ref{Sec_slow_fading},
respectively. The asymptotic analysis of the parameters is given
in Section \ref{Sec_asymptotic}. A nested-lattice based LA-GPC
scheme using the proposed design parameters is illustrated in
Section \ref{Sec_lattice}. Simulation results are given in Section
\ref{Sec_simulation}. Finally, Section \ref{Sec_Conclusion}
concludes this paper.

\vspace{-3mm}
\section{Background and System Model}\label{Sec_system_model}
In this section we first introduce the system model in
consideration. We then review the problems of DPC and LA-GPC in
fading channels. After that we will derive the rate formula of the
fading cognitive channel with LA-GPC.\footnote{In this paper, the
superscript $(.)^{\mathrm{T}}$ and $(.)^{\mathrm{H}}$ denote the
transpose and complex conjugate, respectively. Letters without and
with underlines denote variables and vectors, respectively. If a
letter is Italic capital, the corresponding variable or vector is
random. Boldface capital letters denote deterministic matrices.
$\mathbf{A}(i,\,j)$ denotes the entry at the $i$th row and the
$j$th column of the matrix $\mathbf{A}$. The covariance matrix of
the random vector $[A\,B]^{\mathrm{T}}$ is denoted by
$\Sigma_{A,\,B}$. $|\mathbf{A}|$ and $|a|$ represent the
determinant of the square matrix $\mathbf{A}$ and the absolute
value of the scalar variable $a$, respectively. $\mathbf{I}_n$ is
the identity matrix with dimension $n$. $I(A;B)$ denotes the
mutual information between $A$ and $B$. The trace of $\mathbf{A}$
is denoted by tr($\mathbf{A}$). The diagonal matrix whose diagonal
entries are $\underline{a}$ is denoted by diag($\underline{a}$).
All the logarithm operations are of base 2 such that the unit of
rates is in bit.} \vspace{-3mm}
\subsection{The fading cognitive channel}
\begin{figure}[htp]
\centering
\epsfig{file=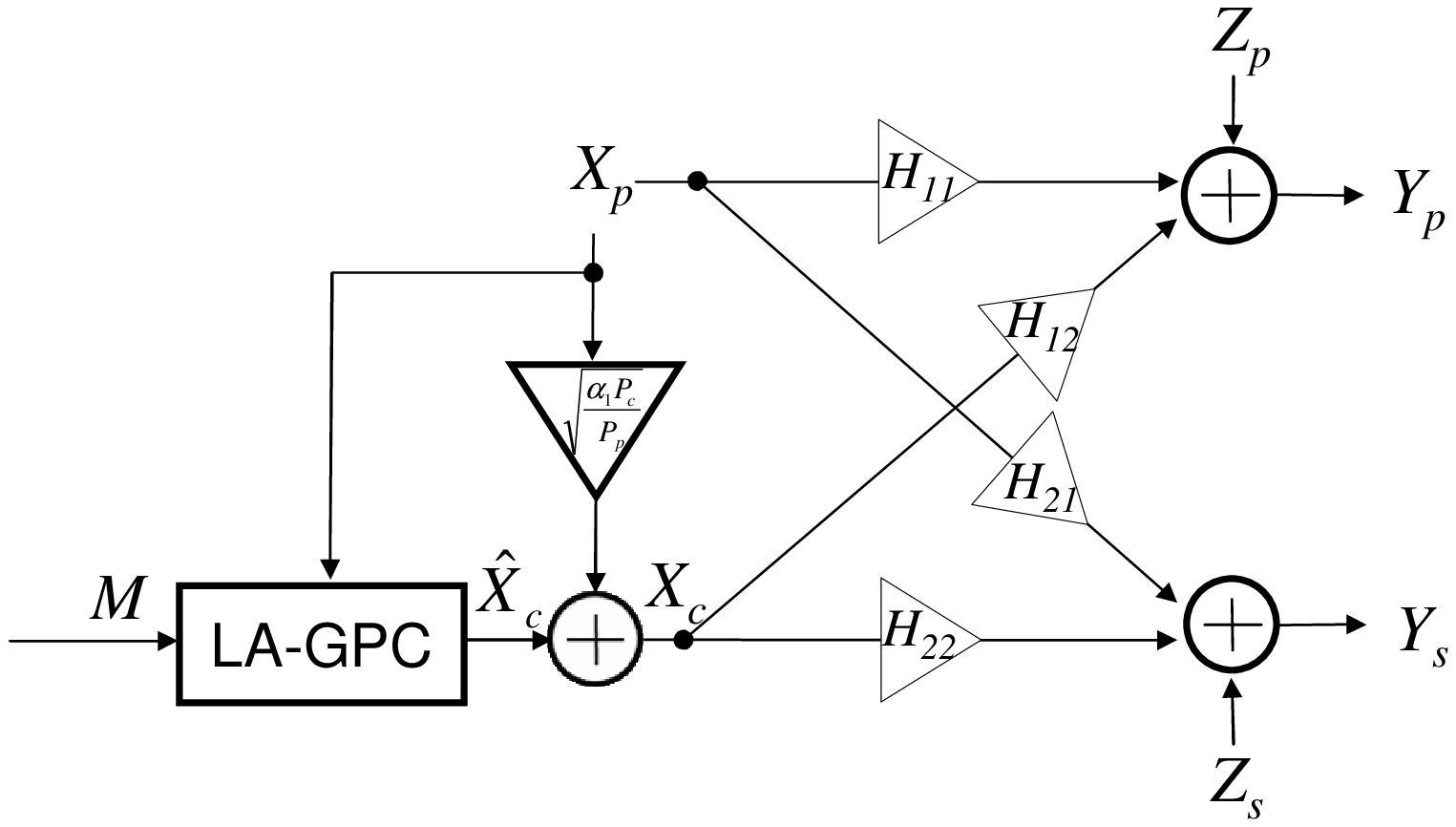, width=0.6\textwidth}
\caption{The model of the cognitive radio channel.}
\label{Fig_system_model}
\end{figure}

\vspace{-2mm} The cognitive channel coined by
\cite{Devroye_CR_achievable_rate} is shown in Fig.
\ref{Fig_system_model}. The channel gains between primary
transmitter and receiver, primary transmitter and CR receiver, CR
transmitter and primary receiver, and CR transmitter and receiver
are denoted by $H_{11},H_{21},H_{12}$, and $H_{22}$, respectively.
The primary and CR transmitted signals are $X_p$ and $X_c$ with
transmit power constraint $P_p$ and $P_c$, respectively. The
unidirectional arrow from the primary transmitter to the CR
transmitter means that the CR transmitter unilaterally knows the
primary user's codewords non-causally. The feasibility of this
assumption is discussed in \cite{Viswanath_CR}. The signals at the
primary and CR receivers are
\begin{equation}
\left[\begin{array}{c} Y_p\\
Y_s\end{array}\right]= \left[\begin{array}{cc} H_{11} &H_{12}\\
H_{21} &H_{22}\end{array}\right]\left[\begin{array}{cc} 1 & 0\\
\sqrt{\frac{\alpha_1P_c}{P_p}}
 &
 1\end{array}\right]\left[\begin{array}{c}X_p\\\hat{X}_c\end{array}\right]+\left[\begin{array}{c} Z_p\\
Z_s\end{array}\right],
\end{equation}
where $\hat{X}_c$ is the CR user's own signal after being
precoded. The CR transmitter relays the primary user's signal to
maintain the performance of the primary link. The relaying ratio
$\alpha_1$ is the percentage of CR's transmit power used for
relaying the primary signal. Thus the factor
$\sqrt{\alpha_1P_c/P_p}$ is to ensure that the relaying power is
$\alpha_1P_c$. The circularly symmetric complex additive white
Gaussian noises (AWGN) at the primary and CR receivers are denoted
by ${Z}_p\sim \mathpzc{CN}(0,\sigma_{Z_p}^2)$ and ${Z}_s\sim
\mathpzc{CN}(0,\sigma_{Z_s}^2)$, respectively.

In the following discussion we assume that the primary receiver knows
$H_{11}$ perfectly, but only the statistics of $H_{11}$ are known
at primary transmitter. Similarly, the CR receiver knows all four
channel gains perfectly, but only the statistics of them are known at
the CR transmitter. We assume that the primary and CR users are subject to
slow or fast fading channels simultaneously. In this paper the
four channels $H_{ij}$, $1 \leq i,j \leq 2$, are assumed to be
Rice distributed with mean and variance denoted by $\mu_{ij}$ and
$\sigma^2_{ij}$, respectively. We also let
$|\mu_{ij}|^2+\sigma_{ij}^2=1$ to keep the calculation of the
received signal-to-noise ratio (SNR) simple.

\vspace{-3mm}
\subsection{LA-GPC in fading channels} \label{sec_LA_intro}
\vspace{-2mm} Since only the statistics of the fading channels are
known at the transmitter, the CR user cannot use the classical DPC
to perfectly cancel the interference due to $X_p$. Instead, the
DPC is replaced by the LA-GPC which is more suitable for this
task. For brevity, the CR user's channel is recast as
\begin{equation} \label{Eq_fading_paper_channel_Random}
Y=H_xX+H_sS+Z,
\end{equation}
with
\begin{align} \label{eq_GPC_Para}
& H_s=H_{21}+\sqrt{\frac{\alpha_1 P_c}{P_p}}H_{22}, \;
H_x=H_{22},\; Y=Y_s,\; X=\hat{X}_c, \; S=X_p, \mbox{ and } Z=Z_s,
\end{align}
where $X$, $S$, and $Z$ are assumed to be independent complex
Gaussian random variables with zero mean and variances (powers)
$P_x$, $P_s$, and $P_z = \sigma_{Z_s}^2$, respectively. The fading
channel gains $H_x$ and $H_s$ of the signal and interference,
respectively, are known perfectly at the receiver, but only their
statistics are known at transmitter. For illustration purpose, we
first focus on the fast fading channel in this subsection. The
capacity of this channel can be modified from
\cite{Gelfand_Pinsker} as
\begin{equation} \label{Eq_GelPin}
\tb{\mathrm{sup}}{\mathpzc{f}(u|s),f(\cdot)} E\{R(H_x,\,H_s)\},
\end{equation}
where
\begin{equation} \label{eq_LA_rate_given_h}
R(H_x,\,H_s)=I(U;Y|H_x,H_s)-I(U;S),
\end{equation}
$U$ is an auxiliary random variable with distribution obtained via
the conditional distribution $\mathpzc{f}(u|s)$, and $f(\cdot)$ is
a deterministic function such that $X=f(U,S)$. We select
$f(U,S)=U-\alpha_2 S$ which makes,
\begin{equation} \label{Eq_LS_Bennatan}
U=X+\alpha_2 S.
\end{equation}
Such strategy function selection is called the \textit{linear
assignment} strategy and $\alpha_2$ is the precoding coefficient.
With perfect CSIT, we can choose $\alpha_2$ as the linear minimum
mean square error (MMSE) filter coefficient $\alpha_c \triangleq
|H_x|^2P_x/(|H_x|^2P_x+P_z)$ to estimate $X$ from the
interference-free channel $Y=H_xX+Z$, and the interference-free
rate is achievable. Costa named the LA-GPC with
$\alpha_2=\alpha_c$ as DPC \cite{CostaDPC}. However, in our
setting, the selection of the filter $\alpha_2$ must rely only on
the channel statistics. Thus the MMSE filter used in Costa's DPC
cannot be used here.

\vspace{-3mm}
\subsection{Achievable rate of CR with LA-GPC}\label{subsec_LAinCR} \vspace{-2mm} Before further
discussing the primary and CR users'
performances in fast and slow fading channels, the rate functions
with given channel gains should be derived first. Assuming that the
linear-assignment strategy is used, we have $U=\hat{X}_c+\alpha_2{X}_p$
from (\ref{eq_GPC_Para}) and (\ref{Eq_LS_Bennatan}).
The mutual information with given channel realizations can be
computed as \cite{Pslin_CR1}
\begin{align}
R(h_{22},h_{21})&\triangleq
I(U;{Y}_s|H_{22}=h_{22},H_{21}=h_{21})-I(U;S)\notag\\
&=\log\left((|h_{22}|^2\sigma_{\hat{x}_c}^2+|h_{21}+\sqrt{\frac{\alpha_1P_c}{P_p}}h_{22}|^2P_p+\sigma_{Z_s}^2)\sigma_{\hat{x}_c}^2\right)-\log(|\Sigma_{U,{Y}_s}|),
\label{Eq_I_of_cg}
\end{align}
where
$\sigma_{\hat{x}_c}^2=(1-\alpha_1)P_c$, and $\Sigma_{U,{Y}_s}$ is
\begin{equation} \label{Eq_Sigma_US}
\Sigma_{U,{Y}_s}=\left(
\begin{array}{cc}
 \sigma_{\hat{x}_c}^2+|\alpha_2|^2P_p  & h_{22}^*\sigma_{\hat{x}_c}^2+\alpha_2(h_{21}^*+h_{22}^*\sqrt{\frac{\alpha_1P_c}{P_p}})P_p \\
h_{22}\sigma_{\hat{x}_c}^2+\alpha_2^*(h_{21}+h_{22}\sqrt{\frac{\alpha_1P_c}{P_p}})P_p   & |h_{22}|^2\sigma_{\hat{x}_c}^2+|h_{21}+h_{22}\sqrt{\frac{\alpha_1P_c}{P_p}}|^2P_p+\sigma_{Z_s}^2 \\
\end{array}
\right).
\end{equation}
Similarly, when the CR user is active, the primary user's rate
with given channel realizations can be computed as
\begin{align}
R(h_{11},h_{12}) & \triangleq
I(X_p;Y_p,H_{11}=h_{11},H_{12}=h_{12})
\notag\\
&=\log\frac{(1-\alpha_1)P_c|h_{12}|^2+|\sqrt{P_p}h_{11}+\sqrt{\alpha_1P_c}h_{12}|^2+\sigma_{Z_p}^2}{(1-\alpha_1)P_c|h_{12}|^2+\sigma_{Z_p}^2}.\label{Eq_primary_rate}
\end{align}

\section{Parameter design for fading channels}
In this section we provide semi-analytical methods for finding the
relaying ratios $\alpha_1$ and precoding coefficients $\alpha_2$
under fast and slow fading channels.

\subsection{Fast fading scenario}\label{Sec_fast_fading}
With full CSIT, the CR transmitter must ensure the primary user's
rate unchanged as in Viswanath's setting \cite{Viswanath_CR}. For
fast fading channels, a meaningful definition of such rate is the
\textit{ergodic capacity} \cite{Book_Tse}. Therefore, we take the
rate functions (\ref{Eq_I_of_cg}) and (\ref{Eq_primary_rate})
averaged over all channel realizations to form the following
design criteria
\begin{align} \label{Eq_EG_Traget}
\mbox{maximize  }& E[R(H_{22},H_{21})]\\
\mbox{subject to  }& E[R(H_{11},H_{12})]\geq
E[\log(1+\frac{|H_{11}|^2P_p}{\sigma_{Z_p}^2})].\label{Eq_EG_Constriant}
\end{align}
Since (\ref{Eq_EG_Constriant}) only depends on $\alpha_1$ but not
$\alpha_2$, we can solve (\ref{Eq_EG_Constriant}) for $\alpha_1$
and then (\ref{Eq_EG_Traget}) for $\alpha_2$ independently. The
key to solving (\ref{Eq_EG_Constriant}) is to transform
(\ref{Eq_primary_rate}) into a \textit{ratio of quadratic form}.
By resorting to the moments of the quadratic form \cite{Mathai} we
can find $\alpha_1$ by solving a simple logarithmic equation
numerically. Thus we first rearrange the terms in
(\ref{Eq_primary_rate}) into the following matrix form. Let
\begin{equation} \label{Eq_mu_sigma_def}
\underline{H}_p\triangleq (H_{11},\,H_{12})^{\mathrm{T}} \sim
\mathpzc{CN}(\mathbf{\mu},\Sigma ),
\end{equation}
$\mathbf{\mu}\triangleq(\mu_{11},\,\mu_{12})^{\mathrm{T}}$, and
$\Sigma\triangleq\mbox{diag}([\sigma_{11}^2\;\sigma_{12}^2])$.
Also let
\begin{equation} \label{Eq_PQ_def}
\mathbf{Q}\triangleq\left(\begin{array}{cc}0&0\\
0& (1-\alpha_1)P_c \end{array}\right) \mbox{ and }
\mathbf{P}\triangleq\left(\begin{array}{cc}P_p&\sqrt{\alpha_1P_cP_p}\\
\sqrt{\alpha_1P_cP_p}& \alpha_1 P_c \end{array}\right).
\end{equation}
Without loss of generality, we let both $\sigma_{Z_p}^2$ and
$\sigma_{Z_s}^2$ be 1. Then the primary user's rate formula
becomes
\begin{equation}\label{Eq_primary_matrix_form}
R(H_{11},\,H_{12})=\log(1+\frac{\underline{H}^{\mathrm{H}}_p\mathbf{P}\underline{H}_p}{\underline{H}^{\mathrm{H}}_p\mathbf{Q}\underline{H}_p+1}).
\end{equation}
Note that the right hand side of (\ref{Eq_EG_Constriant}) is
assumed to be a known constant $R^P_{ergodic}$ for the CR
transmitter. Similarly, after some manipulations the CR user's
rate becomes
\begin{equation}\label{Eq_CR_rate_matrix}
R(H_{22},\,H_{21})=\log\left(\frac{\underline{H}_c^{\mathrm{H}}(\mathbf{P}+\mathbf{Q})\underline{H}_c+1}{\underline{H}_c^{\mathrm{H}}\left(c_0(\mathbf{P}+\mathbf{Q})-\mathbf{D}\right)\underline{H}_c+c}\right)+\log\sigma_{\hat{x}_c}^2,
\end{equation}
where $\underline{H}_c=[H_{22},H_{21}]^{T}$,
$c_0=\sigma_{\hat{x}_c}^2+|\alpha_2|^2P_p$, $c=c_0$, and
\begin{align}
\mathbf{D}=\left(
\begin{array}{cc}
 |\alpha_2|^2P_p^2  & \alpha_2(\sigma_{\hat{x}_c}^2+\alpha_2^*\sqrt{\alpha_1P_cP_p})P_p \\
\alpha_2^*(\sigma_{\hat{x}_c}^2+\alpha_2\sqrt{\alpha_1P_cP_p})P_p   & |\sigma_{\hat{x}_c}^2+\alpha_2^*\sqrt{\alpha_1P_cP_p}|^2 \\
\end{array}\right).\notag
\end{align}
The optimization problem described in (\ref{Eq_EG_Traget}) and
(\ref{Eq_EG_Constriant}) is not convex. Thus we propose methods
described in Theorem \ref{Th_1} and Proposition \ref{Th_2} to
solve it sub-optimally with proofs given in Appendix \ref{App_1}
and \ref{App_2}, respectively.

\begin{theorem}\label{Th_1}
The relaying ratio $\alpha_1$ for the fast fading CR channel can
be found sub-optimally by solving
\begin{equation}\label{Eq_appr_ergo_al}
\log(\frac{1+\mu_{\varepsilon_1}}{1+\mu_{\varepsilon_2}})-\frac{\log
e}{2}\cdot\frac{\sigma_{\varepsilon_1}^2}{(1+\mu_{\varepsilon_1})^2}=R_{ergodic}^P,
\end{equation}
where
\begin{align}
\mu_{\varepsilon_1}=\mathbf{\mu}^H\mathbf{S}\mathbf{\mu}+\mbox{tr}(\Sigma\mathbf{S}),\,\,\sigma_{\varepsilon_1}^2=2\mu^H\mathbf{S}\Sigma\mathbf{S}\mu
+ \mbox{tr}(\Sigma\mathbf{S})^2,\,\,
\mu_{\varepsilon_2}=\mathbf{\mu}^H\mathbf{Q}\mathbf{\mu}+\mbox{tr}(\Sigma\mathbf{Q}),\notag
\end{align}
$\mathbf{S}=\mathbf{P+Q}$, $\mu$ and $\Sigma$ are defined right after (\ref{Eq_mu_sigma_def}), and $\mathbf{P}$ and $\mathbf{Q}$ are defined in (\ref{Eq_PQ_def}), respectively.\\
\end{theorem}

After knowing $\alpha_1$, we can treat (\ref{Eq_EG_Traget}) as an
unconstrained optimization problem. Let $B(\alpha_2)\triangleq
\underline{H}_c^{\mathrm{H}}\left(c_0(\mathbf{P}+\mathbf{Q})-\mathbf{D}\right)\underline{H}_c+c$,
which is a convex function of $\alpha_2$. With the fact that the
logarithm is concave, $R(H_{22},\,H_{21})$ is concave with respect
to $\alpha_2$. Therefore, from \cite[p.209]{Boyd_convex_book} we
know that (\ref{Eq_EG_Traget}) is concave. Thus, we can set the
derivative of $E[R(H_{22},\,H_{21})]$ with respect to the complex
conjugate of $\alpha_2$ to zero to find the extreme value
\begin{align}
\frac{\partial E[R(H_{21},\,H_{22})]}{\partial\alpha_2^*}&=-\frac{\partial}{\partial\alpha_2^*}\int\log B(\alpha_2)f_{\underline{H}_c}d\underline{H}_c=-\int\frac{\partial}{\partial\alpha_2^*}\log B(\alpha_2)f_{\underline{H}_c}d\underline{H}_c=-\int\frac{B'(\alpha_2)}{B(\alpha_2)}f_{\underline{H}_c}d\underline{H}_c\notag\\
&=-E\left[\frac{B'(\alpha_2)}{B(\alpha_2)}\right]=0,\label{Eq_general_moment_of_ratio_of_quadratic_form}
\end{align}
where $f_{\underline{H}_c}$ is the probability density function of
$\underline{H}_c$, and the validity of the interchange of the
integration and differentiation of the second equality is obvious.
Note that solving the \textit{general moment of quadratic form}
\cite{Mathai} in
(\ref{Eq_general_moment_of_ratio_of_quadratic_form}) is
intractable. Thus we resort to finding the suboptimal solution.

\begin{proposition}\label{Th_2}
The precoding coefficient $\alpha_2$ for the fast fading CR
channel can be approximated by
\begin{equation}\label{Eq_alpha2_1}
\alpha_2=\frac{(\mu_{22}^*\mu_{21}+\sqrt{\frac{\alpha_1
P_c}{P_p}})(1-\alpha_1)P_c} {(1-\alpha_1)P_c+1}.
\end{equation}
\end{proposition}

\subsection{Slow fading scenario}\label{Sec_slow_fading}
For the quasi-static slow fading channels, the decoding error
probability cannot be arbitrarily small since the transmitter does
not know the reliable transmission rate with the limited channel
knowledge. In this channel, the outage probability \cite{Book_Tse}
for a certain target rate is more suitable than the Shannon
capacity to measure the performance.

For the CR user with a target rate $R^{CR}$, the outage
probability which is defined as $P(R(H_{21},H_{22})<R^{CR})$
should be minimized. Let $P_{out}^{P}$ and $R^P$ be the primary
user's achievable outage probability and outage capacity,
respectively, in the absence of the CR transmitter. Since the CR
system must ensure that the primary user's outage probability does
not increase, we have the following constrained minimization
problem for the outage probability of the CR user
\begin{align}
\mbox{minimize } & P_{out}^{CR}\notag\\
\mbox{subject to } & P(R(H_{21},H_{22})<R^{CR})\leq
P_{out}^{CR} \label{Eq_CR_Pout_constraint}\\
\,\,\,\,\,\,&P(R(H_{11},H_{21})<R^{P})\leq P_{out}^{P}
.\label{Eq_PR_rate_constraint}
\end{align}
Assume that the CR transmitter knows the primary user's target
outage probability $P_{out}^p$ and target rate $R^P$. Similar to
the ergodic case, $\alpha_1$ and $\alpha_2$ can be solved from
(\ref{Eq_PR_rate_constraint}) and (\ref{Eq_CR_Pout_constraint})
sequentially. To reformulate the complex outage constraint
(\ref{Eq_PR_rate_constraint}) into an analytically solvable one,
we resort to the concept of deterministic approximations of
probability inequalities \cite{Pinter}. In \cite{Pslin_CR1} it was
observed that if the mean of the rate is maximized, the deviation
of the rate will also be increased, which may result in a worse
outage capacity. As a result, if we want to achieve the outage
probability constraint efficiently, we need to consider both the
mean and deviation of the rate simultaneously. In addition, to
take the tightness of the approximation into consideration, we
adopt the modified Cantelli's inequality \cite{modified_Cantelli}.
The results are summarized as the following with proofs in
Appendix \ref{App_3} and \ref{App_4}, respectively. The parameters
$\mu_{\varepsilon_2},\, \Sigma,\,\mathbf{P}$, and $\mathbf{Q}$ are
defined in Theorem \ref{Th_1}. With
\begin{equation}\label{Eq_delta1}
\bigtriangleup_1\triangleq\frac{\underline{H}^H_p\mathbf{Q}\underline{H}_p+1}{\underline{H}^H_p\mathbf{P}\underline{H}_p},
\end{equation}
we have

\begin{theorem}\label{Th_3}
The relaying ratio $\alpha_1$ for the slow fading CR channel can
be found by solving the following equation
\begin{equation}\label{Eq_cantelli_approximation}
\mu_{\bigtriangleup_1}+\sqrt{\frac{r}{P_{out}^{P}}-1}\sigma_{\bigtriangleup_1}
= \frac{1}{2^{R^P}-1},
\end{equation}
where $\mu_{\bigtriangleup_1}$ and $\sigma_{\bigtriangleup_1}$ are
the mean and the standard deviation of $\bigtriangleup_1$,
respectively. The constant $r$ can be 1 or 2/9 depending on the
$K$-factor.
\end{theorem}

Since (\ref{Eq_delta1}) is a ratio of quadratic form, we can
resort to \cite{Paollela_ratio_of_quadratic_form} to efficiently
compute the mean and the variance of $\bigtriangleup_1$ by the
following proposition. Then (\ref{Eq_cantelli_approximation}) can
be easily solved by a polynomial equation of $\alpha_1$.

\begin{proposition}\label{Prop_2}
The mean $\mu_{\bigtriangleup_1}$ and variance
$\sigma_{\bigtriangleup_1}^2$ in Theorem \ref{Th_3} are
\begin{equation} \label{Eq_outage_alpha1_mean}
\frac{a}{\mu_{\varepsilon_2}}\left(1-\frac{m}{a\mu_{\varepsilon_2}}+\frac{s}{\mu_{\varepsilon_2}^2}\right)
\mbox{ and  }
\frac{a^2}{\mu_{\varepsilon_2}^2}\left(\frac{t}{a^2}+\frac{s}{\mu_{\varepsilon_2}^2}-\frac{2m}{a\mu_{\varepsilon_2}}\right),
\end{equation}
respectively, where $ a \triangleq \mu^{\mathrm{H}}\mathbf{P}\mu +
\mbox{tr}(\Sigma\mathbf{P}),\, s \triangleq
4\mu^{\mathrm{H}}\mathbf{P}\Sigma\mathbf{P}\mu +
2\mbox{tr}(\Sigma\mathbf{P})^2,\, t \triangleq
4\mu^{\mathrm{H}}\mathbf{Q}\Sigma\mathbf{Q}\mu +
2\mbox{tr}(\Sigma\mathbf{Q})^2,$ and\\ $ m  \triangleq
4\mu^{\mathrm{H}}\mathbf{Q}\Sigma\mathbf{P}\mu +
2\mbox{tr}(\mathbf{Q}\Sigma\mathbf{P}\Sigma)$.
\end{proposition}

Similarly, directly finding the optimal $\alpha_2$ from
calculating the outage probability in
(\ref{Eq_CR_Pout_constraint}) is intractable. The main idea of the
proposed method to find the semi-analytical solution of $\alpha_2$
is described as follows. We first rearrange CR's rate formula
$R(H_{21},H_{22})$ into a weighted sum of non-central chi-square
random variables which can be further approximated by a single
chi-square random variable. Then we can semi-analytically find
$\alpha_2$ minimizing the outage probability $P^{CR}_{out}$ with
the given outage capacity $R^{CR}$. This is summarized in the
following theorem. Note that both $v(\alpha_2)$ and $w(\alpha_2)$
are simple rational functions of $\alpha_2$.

\begin{theorem} \label{Thereom_slow_fading alpha_2}
The precoding coefficient $\alpha_2$ for the slow fading CR
channel can be found by solving
\begin{equation}\label{Eq_alpha2_bound}
\min_{\alpha_2}\,\,\frac{\gamma(\frac{w(\alpha_2)}{2},\frac{cd-1}{2v(\alpha_2)})}{\Gamma(\frac{w(\alpha_2)}{2})},
\end{equation}
where $d=(2^{R^{CR}}-1)/\sigma_{\hat{x}_c}^2$, $\Gamma$ is the
gamma function, $\gamma(.,.)$ is the incomplete gamma
function, and
\begin{align} \label{Eq_vw_outage}
v(\alpha_2)=\frac{1}{2}\frac{\mbox{tr}(\Sigma\mathbf{E}\Sigma\mathbf{E})+2\mu^{\mathrm{H}}\mathbf{E}\Sigma\mathbf{E}\mu}{\mbox{tr}(\Sigma\mathbf{E})+\mu^{\mathrm{H}}\mathbf{E}\mu},\,
w(\alpha_2)=2\frac{\left(\mbox{tr}(\Sigma\mathbf{E})+\mu^{\mathrm{H}}\mathbf{E}\mu\right)^2}{\mbox{tr}(\Sigma\mathbf{E}\Sigma\mathbf{E})+2\mu^{\mathrm{H}}\mathbf{E}\Sigma\mathbf{E}\mu},
\end{align}
where $\mathbf{E}=(1-c_0d)(\mathbf{P}+\mathbf{Q}) +d\mathbf{D}$,
and $c,c_0$ are defined right after (\ref{Eq_CR_rate_matrix}).
\end{theorem}

The problem of solving (\ref{Eq_alpha2_bound}) can be further
approximated by the following corollary.

\begin{corollary}\label{Corollary_1}
The precoding coefficient $\alpha_2$ for the slow fading CR
channel can be approximately found by solving
\begin{equation}\label{Eq_alpha22}
\min_{\alpha_2}\,\,\left(1-\exp\left(-\frac{s(cd-1)}{2v(\alpha_2)}\right)\right)^{\frac{w(\alpha_2)}{2}},
\end{equation}
where
\begin{equation}\label{Eq_sa}
s=\left\{\begin{array}{ll} 1,\,&0<w(\alpha_2)<2\\
(\Gamma(1+w(\alpha_2)/2))^{\frac{-2}{w(\alpha_2)}},\,&w(\alpha_2)>2.
\end{array}\right.
\end{equation}
\end{corollary}

Note that Corollary \ref{Corollary_1} does not need the
computation of the incomplete gamma function in
(\ref{Eq_alpha2_bound}), thus the search of $\alpha_2$ becomes
simpler.

\vspace{-3mm}
\section{Asymptotic analysis}\label{Sec_asymptotic}
Since a Rician fading channel becomes non-fading when its
$K$-factor approaches infinity, and the relaying ratio and
precoding coefficient are well known for non-fading channels
\cite{Viswanath_CR}, we can verify the correctness of the proposed
methods by making the $K$-factor infinity.

\vspace{-3mm}
\subsection{Asymptotic analysis of $\alpha_1$} \vspace{-2mm}
In the high $K$-factor region, the channel gains are almost
deterministic and knowing the mean values of the channels is
almost the same as knowing the channel gains. With the assumption
that $|\mu_{ij}|^2+\sigma_{ij}^2=1,\,1\leq i,j\,\leq 2$,
parameters and performances of slow and fast fading channels
should converge to that of the non-fading channel with unity
channel gain. Thus both $\sigma_{\varepsilon_1}$ and
$\sigma_{\bigtriangleup_1}$ approach zero. Besides,
\[
\lim_{K\rightarrow\infty}\mu_{\varepsilon_1}=\varepsilon_1,\,\lim_{K\rightarrow\infty}\mu_{\varepsilon_2}=\varepsilon_2,
\]
where the definitions of $\varepsilon_1$ and $\varepsilon_2$ are
given right after (\ref{Eq_Appr_ergo_R}). For the fast fading
case, the left hand side of (\ref{Eq_appr_ergo_al}) becomes
\begin{equation}\label{Eq_asymp_a1}
\lim_{K\rightarrow\infty}E[R(H_{11},H_{12})]=\log(\frac{1+\mu_{\varepsilon_1}}{1+\mu_{\varepsilon_2}})=\log(\frac{1+\varepsilon_1}{1+\varepsilon_2})=\log(1+\bigtriangleup_1^{-1})=\log(1+P_p),
\end{equation}
where $\bigtriangleup_1$ is defined in (\ref{Eq_delta1}). On the
other hand, for the slow fading case
(\ref{Eq_cantelli_approximation}) becomes
\begin{equation}\label{Eq_asymp_a12}
\lim_{K\rightarrow\infty}
\left(\mu_{\bigtriangleup_1}+\sqrt{\frac{r}{P_{out}^{P}}-1}\sigma_{\bigtriangleup_1}\right)=\lim_{K\rightarrow\infty}\mu_{\bigtriangleup_1}=\bigtriangleup_1=P_p^{-1}.
\end{equation}
From (\ref{Eq_asymp_a1}) and (\ref{Eq_asymp_a12}) we can find that
$\alpha_1$ converges in both fast and slow fading scenarios to the
same value. From the definition of $\bigtriangleup_1$ we have
\[
\frac{P_p+\alpha_1P_c+2\sqrt{\alpha_1P_cP_p}}{(1-\alpha_1)P_c+1}=P_p,
\]
which results in the same $\alpha_1$ as that in the non-fading CR case
\cite{Viswanath_CR}. Thus we can conclude that the $\alpha_1$'s found by the proposed
methods in both fast and slow fading
channels converge to that of the non-fading case when
$K\rightarrow\infty$.

\vspace{-3mm}
\subsection{Asymptotic analysis of $\alpha_2$} \vspace{-2mm}
Again the precoding coefficient $\alpha_2$ should coincide with
that of the non-fading case when $K \rightarrow \infty$. In the
following we verify this statement.

\subsubsection{Fast fading} When $K \rightarrow \infty$,
both $\mu_{22}$ and $\mu_{21}$ approach 1, and
(\ref{Eq_alpha2_1}) becomes
\begin{equation}\label{Eq_asymp_a2_fast}
\lim_{K\rightarrow\infty}\alpha_2=\frac{(1+\sqrt{\frac{\alpha_1
P_c}{P_p}})(1-\alpha_1)P_c}
{(1-\alpha_1)P_c+1}=(1+\sqrt{\frac{\alpha_1
P_c}{P_p}})\alpha_2^{MMSE},
\end{equation}
where $\sqrt{\frac{\alpha_1 P_c}{P_p}}$ is due to the CR's
relaying of the primary signal and $\alpha_2^{MMSE}$ is the
precoding coefficient of the non-fading case \cite{Viswanath_CR}.
Thus it is verified that the proposed method also results in the MMSE
linear assignment as the perfect CSIT case when $K \rightarrow \infty$.

\subsubsection{Slow fading}\label{Sec_Asymp_alpha2_slow}The
asymptotic property of $\alpha_2$ is described in the following
corollary with proof given in Appendix \ref{App_5}.
\begin{corollary}\label{L_1}
When $K$-factor $\rightarrow\infty$, $\alpha_2$ derived in Theorem
\ref{Thereom_slow_fading alpha_2} converges to
$(1+\sqrt{\frac{\alpha_1 P_c}{P_p}})\alpha_2^{MMSE}.$\\
\end{corollary}
From (\ref{Eq_asymp_a2_fast}) and Corollary \ref{L_1}, we conclude
that $\alpha_2$'s for slow and fast fading channels converge to
the same value in \cite{Viswanath_CR} when the $K$-factor
$\rightarrow\infty$.

\vspace{-3mm}
\section{Practical lattice-based linear-assignment Gel'fand-Pinsker coding} \label{Sec_lattice}
All the previous results are based on the LA-GPC with theoretical,
unstructured Gaussian random codebooks. In the following we
introduce a practical lattice-based coding to implement the LA-GPC
for the CR user. With the carefully designed filters specified in
Theorem \ref{theorem_nested_DPC_LA}, the proposed lattice-based
LA-GPC is a non-trivial extension of the well-known lattice-based
DPC \cite{ShamaiMultibinning}. First, the side-information filter
is selected according to the aforementioned precoding coefficient
$\alpha_2$ instead of directly using the MMSE filter $\alpha_{c}$
described after (\ref{Eq_LS_Bennatan}) for the interference-free
channel as in Section \ref{sec_LA_intro} of
\cite{ShamaiMultibinning}. Secondly, with only the statistics of
the CSIT, the receiver filter selection is more involved than
simply choosing it as $\alpha_{c}$ in DPC
\cite{ShamaiMultibinning}\cite{MIMOP2P_El_Gamal_IT04}.
Specifically, this filter must be different from the
side-information filter, and this is contrary to the common
practice in the lattice-based DPC
\cite{ShamaiMultibinning}\cite{MIMOP2P_El_Gamal_IT04}. Finally, we
adopt the finite dimensional self-similar lattices to implement
the lattice coding structure which is much more feasible than the
lattice-based DPC in \cite{ShamaiMultibinning} using a very long
codeword length. The proposed coding works very well with a
reasonable codeword length (and decoding latency), and our
simulation in Section \ref{sec_lattice_simu} verifies this.

To illustrate the lattice-based LA-GPC, we focus on the
following channel corresponding to
(\ref{Eq_fading_paper_channel_Random}) with discrete time index
$t$ as
\begin{equation} \label{Eq_fading_paper_channel_t}
Y(t)=H_x(t) X_L(t) + H_s(t) S(t)+ Z(t),
\end{equation}
where $1 \leq t \leq T$, $T$ is the codeword length. To emphasize
the difference between lattice coding and random Gaussian
codebooks used in the previous sections, we use $X_L(t)$ instead
of $X(t)$ to represent the lattice-coded signal. As in Section
\ref{subsec_LAinCR}, we first focus on the rate
(\ref{eq_LA_rate_given_h}) with given channel realizations
$H_x(t)=h_x(t)$ and $H_s(t)=h_s(t)$. We also rewrite
(\ref{Eq_fading_paper_channel_t}) in an equivalent real super
channel to present the lattice coding more easily. By
concatenating all $T$ symbols, the channel becomes
\begin{equation} \label{eq_GDPC_color_channel}
\underline{Y}=\mathbf{H_x} \underline{X}_L+\mathbf{H_s}
\underline{S}+\underline{Z},
\end{equation}
where $\underline{X}_L=(\underline{X}_{L,1}^{\mathrm{T}},\ldots,
\underline{X}_{L,\mathrm{T}}^{\mathrm{T}})^{\mathrm{T}}$ and
$\underline{X}_{L,t}=(\mathrm{Re}\{X_L(t)\},\mathrm{Im}\{X_L(t)\})^{\mathrm{T}}$.
The non-causally known side-information vector at the transmitter
$\underline{S}$ and the noise vector $\underline{Z}$ are obtained
similarly from $S(t)$ and $Z(t)$, respectively, as
$\underline{X}_L$ from $X_L(t)$. The covariance matrices of
$\underline{S}$ and $\underline{Z}$ are denoted by $\Sigma_{s}$
and $\Sigma_{z}$, respectively. The $2T \times 2T$ block-diagonal
real channel matrix $\mathbf{H_x}$ is
$diag([\mathbf{H_x}^1\cdots\mathbf{H_x}^{T}])$, where the $t$th
diagonal term is
\[
\mathbf{H_x}^t=\left[
  \begin{array}{cc}
   \mathrm{Re}\{h_x(t)\} & -\mathrm{Im}\{h_x(t)\} \\
   \mathrm{Im}\{h_x(t)\} & \mathrm{Re}\{h_x(t)\}
  \end{array}
\right],
\]
and $\mathbf{H_s}$ is formed from $H_s(t)$ in the same way as
$\mathbf{H_x}$ from $H_x(t)$. The channel input power constraint
is $(1-\alpha_1)P_c/2$ because the CR user transmits its own signal with power
$P_c/2$ per real dimension, and relays
the primary user's signal with power $\alpha_1P_c/2$.

We will first give a brief review of the lattice codebook, then
introduce our proposed lattice coding. A $2T$-dimension real
lattice $\Lambda$ is defined as $\Lambda=\{ \mathbf{Gb : b} \in
\mathds{Z}^{2T}\}$, where $\mathbf{G}$ is the $2T \times 2T$
generator matrix of $\Lambda$. The Voronoi region $\mathpzc{V}$ is
the set of points $\mathbf{g} \in \mathds{R}^{2T}$ which are
closest to $\mathbf{0}$ in Euclidean distance than to other
lattice points $\lambda \in \Lambda$. Every $\mathbf{g \in
\mathds{R}}^{2T}$ can be uniquely written as
$\mathbf{g}=\lambda+\mathbf{n}_g$, where $\lambda \in \Lambda$ and
$\mathbf{n}_g \in \mathpzc{V}$. With quantizer input $\mathbf{g}$,
the lattice quantizer associated with $\mathpzc{V}$ is defined as
$ Q(\mathbf{g})=\lambda,\;\;\mathrm{if} \; {\mathbf{g}\in
\lambda+\mathpzc{V}}. $ The modulo-$\Lambda$ operation associated
with $\mathpzc{V}$ is then
\begin{equation} \label{Eq_def_mod}
\mathbf{g} \; \mathrm{mod} \; \Lambda = \mathbf{g}-Q(\mathbf{g}).
\end{equation}
We define the nested-lattice codes with length $2T$ as
\begin{definition}
Let $\Lambda_c$ be a lattice and $\Lambda_q$ be a sublattice of
it, that is, $\Lambda_q \subseteq \Lambda_c$. The codeword set of
the nested lattice code is $ \mathpzc{C}_c = \{ \Lambda_c \;
\mathrm{mod} \; \Lambda_q \} \triangleq \{ \Lambda_c \cap
\mathpzc{V}_q \}. $ The code rate of this nested lattice code
is $
  R_c=\frac{1}{T} \log ||\mathpzc{V}_q||/||\mathpzc{V}_c||,
$ where $\mathpzc{V}_c$ and $\mathpzc{V}_q$ are the fundamental
Voronoi regions of $\Lambda_c$ and $\Lambda_q$, respectively. And
for a bounded Jordan-measurable region $\mathpzc{V}\subset$
$\mathbb{R}^{2T}$, $||\mathpzc{V}||$ denotes the volume of
$\mathpzc{V}$.
\end{definition}

For conciseness of the paper, we focus on the slow fading channels
where the channels remain constant within a codeword length. The proposed
scheme can be easily modified for the fast fading channels
\cite{scFadingDPC}. The proposed coding works as follows
\\
\noindent \textbf{Encoder:} The encoder selects a codeword
$\mathbf{c_c} \in \mathpzc{C}_c$ according to the message index
and sends
\begin{equation} \label{Eq_2_nestedX}
\underline{X}_L=\sqrt{(1-\alpha_1)P_c}\left(\mathbf{(c_c-F_s}\underline{S}-\underline{D})
\; \mathrm{mod} \; \Lambda_q\right),
\end{equation}
where the $2T \times 2T$ side-information filter $\mathbf{F_s}$ is
formed from $\alpha_2$ in Theorem \ref{Thereom_slow_fading
alpha_2} as
\begin{equation} \label{Eq_WB}
\mathbf{I}_T \otimes \left[\begin{array}{cc}
\mathrm{Re}\{\alpha_2\}&\mathrm{-Im}\{\alpha_2\}\\
\mathrm{Im}\{\alpha_2\}&\mathrm{Re}\{\alpha_2\}
\end{array}\right],
\end{equation}
and $\otimes$ denotes the Kronecker product. The dither signal
$\underline{D}$, uniformly distributed in $\mathpzc{V}_q$ and
independent of the channel, is known to both the transmitter and
the receiver. This dither plays a critical role in making
$\underline{X}_L$ independent of $\mathbf{c_c}$ and
$\underline{S}$ \cite{scFadingDPC}. The second moment
$P(\mathpzc{V}_q)$ \cite{MIMOP2P_El_Gamal_IT04} of $\Lambda_q$ is
set $1/2$ to satisfy the power constraint $(1-\alpha_1)P_c/2$.
\\
\noindent \textbf{Decoder:}  The decoder performs the following
operation on $\underline{Y}$
\begin{equation} \label{Eq_2_nested_y_cap}
\underline{\hat{Y}} = \mathbf{L(F_r}\underline{Y}+\underline{D}).
\end{equation}
After some manipulations \cite{scFadingDPC}, $\underline{\hat{Y}}$
in (\ref{Eq_2_nested_y_cap}) can be rewritten as
\begin{equation} \label{Eq_2_nested_y_cap2}
\underline{\hat{Y}} = \mathbf{L}(\mathbf{c'_c})+\mathbf{e},
\end{equation}
where $\mathbf{c'_c} \in \Lambda_q + \mathbf{c_c} \subset
\Lambda_c$ and
\begin{equation} \label{Eq_LA_e}
\mathbf{e \triangleq
(F_r\tilde{H}-I}_{2T})\underline{D}+\mathbf{(F_rH_s-F_s)}\underline{S}+\mathbf{F_r}\underline{Z},
\end{equation}
where $\mathbf{\tilde{H}}=\sqrt{(1-\alpha_1)P_c}\mathbf{H}_x$. The
receiver filter $\mathbf{F_r}$ is obtained according to the MMSE
filter for estimating the auxiliary random variable $U$ in
(\ref{Eq_LS_Bennatan}) from $Y$ in
(\ref{Eq_fading_paper_channel_Random}) as
\cite{scFadingDPC}\cite{scThesis}
\begin{equation} \label{Eq_Wr}
\mathbf{F_r}=(\frac{1}{2}\mathbf{\tilde{H}^{\mathrm{T}}}+\mathbf{W}\Sigma_s\mathbf{H_s}^{\mathrm{T}})(\frac{1}{2}\mathbf{\tilde{H}\tilde{H}^{\mathrm{T}}}+\mathbf{H_s}\Sigma_s\mathbf{H_s}^{\mathrm{T}}+\Sigma_z)^{-1},
\end{equation}
where
$\Sigma_s$ is the covariance matrix of $\underline{S}$. The
whitening matrix filter $\mathbf{L}$ for $\mathbf{e}$ must satisfy
\begin{equation} \label{Eq_L}
\mathbf{L}^{\mathrm{T}}\mathbf{L}=\Sigma^{-1}_E,
\end{equation}
where $\Sigma_E$ is the covariance matrix of $\mathbf{e}$ when $T
\rightarrow \infty$ as
\begin{equation} \label{Eq_Sigma_E}
\Sigma_E=\frac{1}{2}(\mathbf{F_r}
\tilde{\mathbf{H}}-\mathbf{I}_{2T})(\mathbf{F_r}
\tilde{\mathbf{H}}-\mathbf{I}_{2T})^{\mathrm{T}}+
(\mathbf{F_rH_s-F_s}) \Sigma_s (\mathbf{F_rH_s-F_s})^{\mathrm{T}}
+ \mathbf{F_r} \Sigma_z \mathbf{F_r}^{\mathrm{T}}.
\end{equation}

Then we can use the generalized minimum Euclidean distance lattice
decoder to decode $\mathbf{c_c}$. First the decoder finds
\begin{equation} \label{Eq_2_minEuldecoder}
\hat{\mathbf{b}}=\tbsmall{ \mathrm{arg} \; \mathrm{min}}{
\mathbf{b} \in \mathds{Z}^{2T}}
|\mathbf{\hat{y}-LG}_c\mathbf{b}|^2,
\end{equation}
where $\mathbf{G}_c$ is the generator matrix of the channel coding
lattice $\Lambda_c$. After that the decoded codeword is $
\mathbf{\hat{c}}_c=[\mathbf{G}_c \mathbf{\hat{b}}] \; \mathrm{mod}
\; \Lambda_q. $

With ``good nested lattices'' defined in
\cite{MIMOP2P_El_Gamal_IT04}, we have the following result. The
proof is omitted and can be found in \cite{scFadingDPC}.

\begin{theorem} \label{theorem_nested_DPC_LA}
Let the side-information filter $\mathbf{F_s}$, receiver filter
$\mathbf{F_r}$, and whitening filter $\mathbf{L}$ be selected as
(\ref{Eq_WB}), (\ref{Eq_Wr}), and (\ref{Eq_L}), respectively.
Based on sequences of good nested lattices, the coding specified
in (\ref{Eq_2_nestedX})-(\ref{Eq_2_minEuldecoder}) with
(\ref{eq_GPC_Para}) is able to achieve the linear-assignment rate
$R(h_{22},h_{21})=\frac{1}{T}\log|\Sigma_E|^{-1}$ under power
constraint $(1-\alpha_1)P_c$ when $T \rightarrow \infty$ where
$\Sigma_E$ is defined in (\ref{Eq_Sigma_E}).
\end{theorem}

It can be proved that with full CSIT, the proposed coding reduces
to the DPC in
\cite{ShamaiMultibinning}\cite{MIMOP2P_El_Gamal_IT04}. And
according to our simulation, the DPC suffers severely when it is naively
applied to channels without perfect CSIT.

\vspace{-3mm}
\section{Simulation results} \label{Sec_simulation}
In this section we demonstrate the performance of the proposed
LA-GPC based CR system under both fast and slow fading channels.
We also show the performance of the proposed practical lattice
encoder/decoder which implements the LA-GPC. For simplicity, the
four channels in Fig. \ref{Fig_system_model} are assumed
independent Rician with the same $K$-factors.

\vspace{-3mm}
\subsection{Fast fading} 
\begin{figure}[htp]
\centering
\epsfig{file=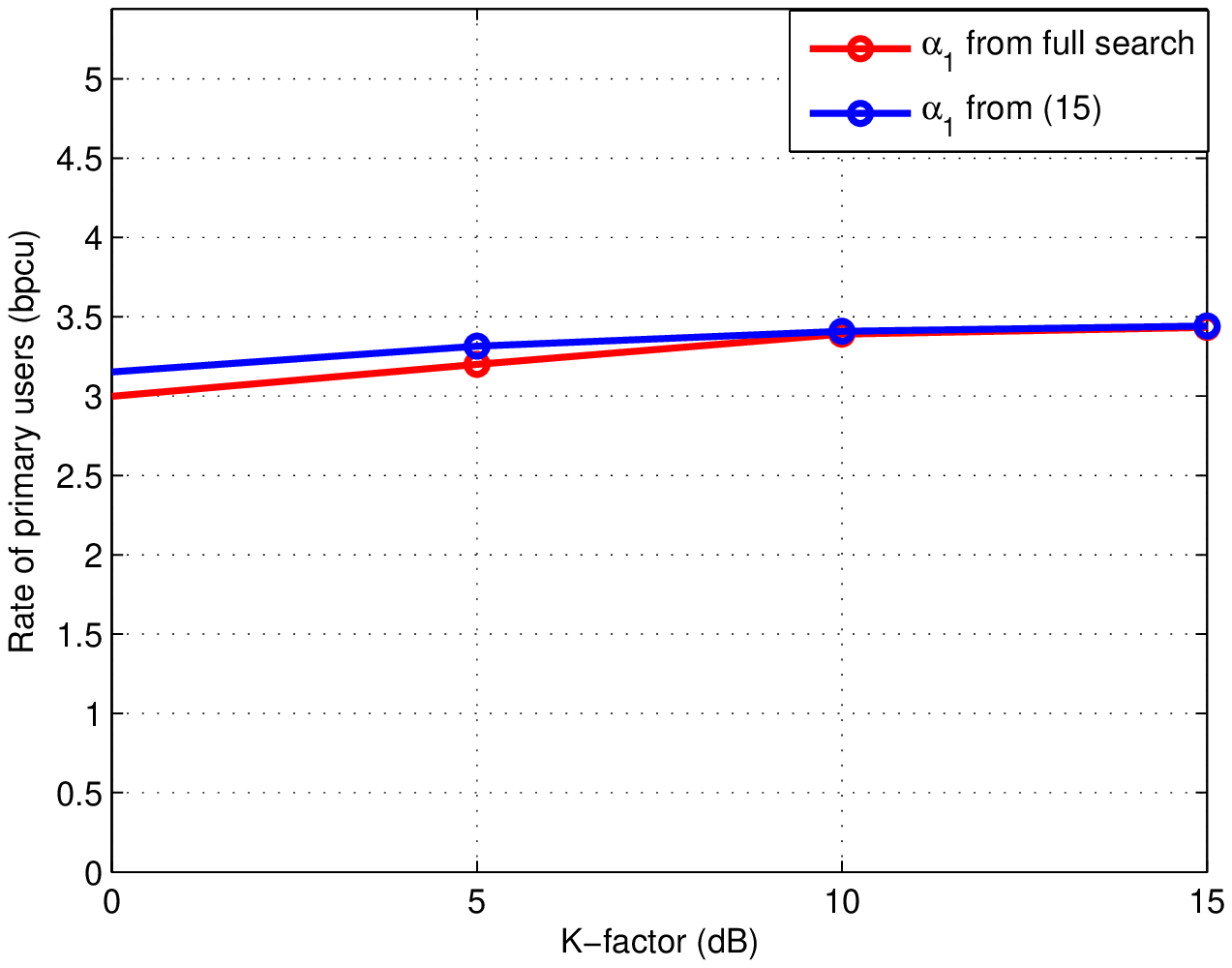, width=0.6\textwidth}
\caption{Comparison of the primary user's ergodic capacities under
full search and the proposed method with $P_c=P_p=10$ and
$\sigma_{Z_s}^2=\sigma_{Z_p}^2=1$.} \label{Fig_ergo}
\end{figure}

\begin{figure}[htp]
\centering \epsfig{file=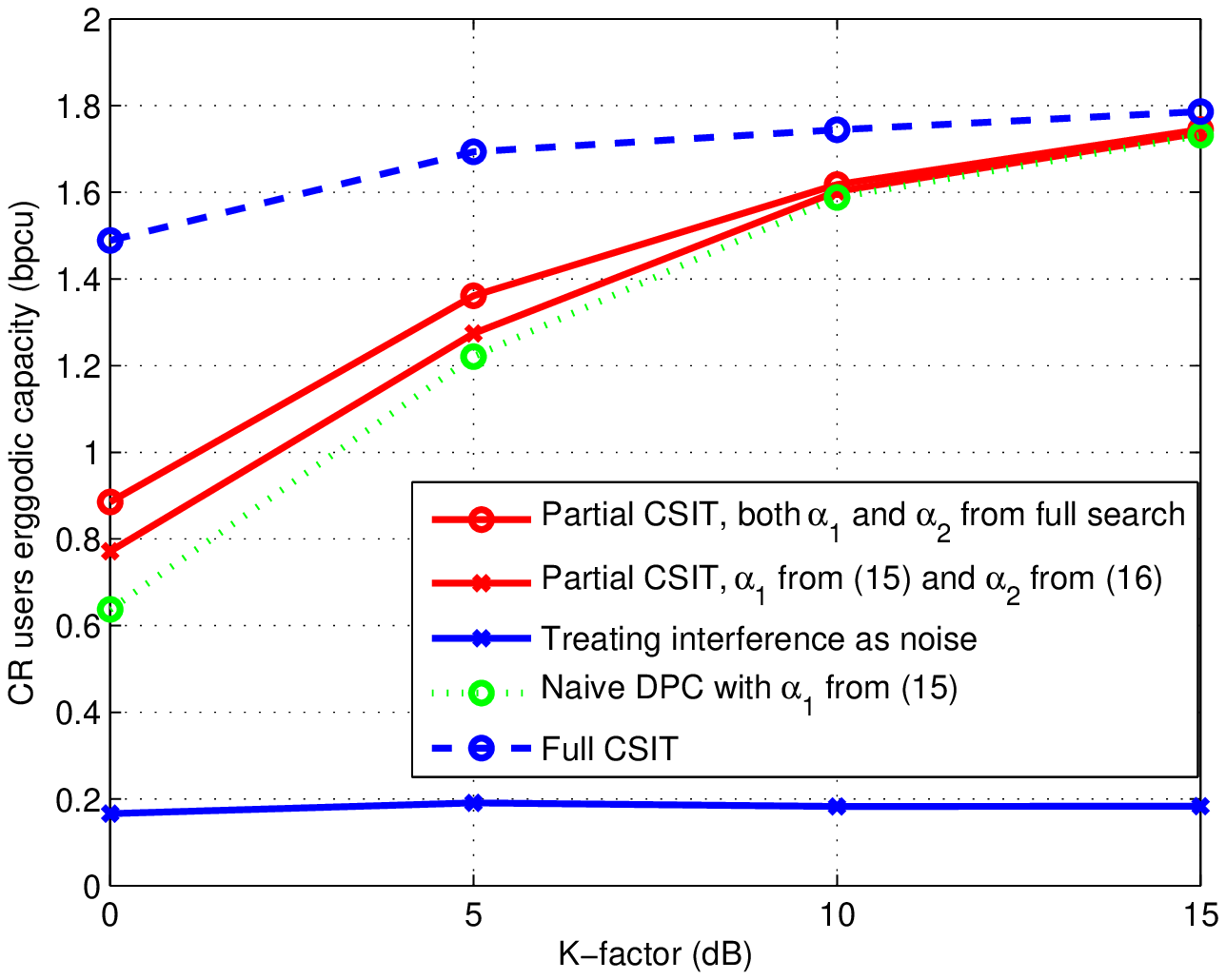,
width=0.6\textwidth} \caption{Comparison of the CR user's ergodic
capacities under full search and the proposed method with
$P_c=P_p=10$ and $\sigma_{Z_s}^2=\sigma_{Z_p}^2=1$.}
\label{Fig_ergo_CR}
\end{figure}

\vspace{-2mm}
To verify the goodness of the proposed approximation, we first
compare it with brute-force full search in terms of the primary
user's rate. Signal and noise powers are set as $P_c=P_p=10$ and
$\sigma_{Z_p}^2=\sigma_{Z_s}^2=1$. This comparison is shown in
Fig. \ref{Fig_ergo}. The unit of the vertical axis is bits per
channel use, bpcu. We can see that the proposed approximation
performs well for all $K$-factors. That is to say, the primary
user's rate is approximately the same as that with $\alpha_1$
obtained by brute-force search. More specifically, the proposed
method may over-design $\alpha_1$ slightly at the small $K$-factor
region such that the primary user's rate becomes larger than the
target. At the large $K$-factor region, $\alpha_1$ obtained by the
proposed method almost coincides with that obtained by full
search. The CR user's ergodic capacities using $\alpha_1$ and
$\alpha_2$ from the proposed method is shown in Fig.
\ref{Fig_ergo_CR}. It can be observed that the over-designed
$\alpha_1$ slightly increases the primary's rate at the cost of
the CR user's rate. However, since $\alpha_1$ is only slightly
over-designed, the resulting change of the CR user's rate is
small. In addition, we also compare cases of treating interference
as noise and with full CSIT. The former neglects any useful
information to cancel the interference, and incurs a huge rate
loss for all $K$-factors considered. The ergodic capacities for
the partial and full CSIT cases converge when the $K$-factor is
large. This phenomenon has been verified in Section
\ref{Sec_asymptotic}. The rate loss at the low $K$-factor region
is due to two factors. One is the imperfect CSIT. The other is
that the linear assignment strategy may not be optimal for the
partial CSIT case. Until now, the optimal strategy for the partial
CSIT case is still an open problem. Note that all the LA-GPC
curves can also be obtained from the lattice-code achievable rate
$1/T\log|\Sigma_E|^{-1}$ with filters specified in Theorem
\ref{theorem_nested_DPC_LA} and $T\rightarrow \infty$. The naive
DPC scheme assumes that the channels always take values at their
means, and uses these mean values to design the DPC filter. That
is, the $\mathbf{F_s}$ filter is naively chosen according to
$\alpha_c$ in Section \ref{sec_LA_intro} with $H_x$ replaced by
$E[H_x]$, and $\mathbf{F_r}=\mathbf{F_s}\mathbf{H_x}^{-1}$.
It can be observed that the naive DPC scheme has a significant
rate loss compared to the proposed method in a wide range of
K-factors.
As expected, the ergodic capacities
of all cases except treating interference as noise converge to the
same value when the $K$-factor is large.

\vspace{-3mm}
\subsection{Slow fading} 
\begin{figure}[htp]
\centering
\epsfig{file=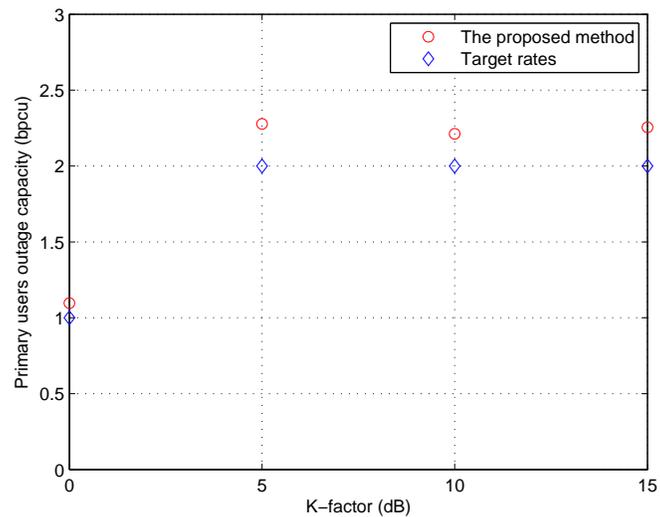,
width=0.6\textwidth} \caption{Comparison of the primary user's
outage capacities obtained by the proposed method and the target
rates with $P_c=P_p=10$ and $\sigma_{Z_s}^2=\sigma_{Z_p}^2=1$.}
\label{Fig_outage_alpha1}
\end{figure}

\vspace{-2mm}
We first demonstrate the effectiveness of the approximation for
$\alpha_1$ in terms of the primary user's outage capacity. A good
$\alpha_1$ will make $P(R(H_{11},H_{21})<R^P)$ as close to
$P_{out}^P$ as possible. We then compare the values of $\alpha_2$
obtained from (\ref{Eq_alpha2_bound}) and full search. Finally, we
illustrate the CR user's outage probabilities with $\alpha_2$'s
from these two methods. Signal and noise powers are set as
$P_c=P_p=10$ and $\sigma_{Z_p}^2=\sigma_{Z_s}^2=1$. Since
different channel conditions can support different outage capacity
and outage probability pairs, it is more reasonable to consider
the supportable pairs than using the same pair for different
$K$-factors. Thus, in the following, we set primary user's outage
capacity (in bpcu) and probability pairs as (1, 0.1), (2, 0.1),
(2, 0.01), and (2, 0.01) for $K$-factors of 0, 5, 10, and 15 dB,
respectively. Moreover, the CR user's target rates are set as 0.2,
0.5, 1, and 1.5 bpcu, respectively. Fig. \ref{Fig_outage_alpha1}
checks whether the primary user's performance constraint
(\ref{Eq_PR_rate_constraint}) is satisfied. Compared to the
baseline case without the presence of the CR transmitter, we find
that the approximation of $\alpha_1$ is very tight in the low
$K$-factor region, and is good for medium and high $K$-factors.
The tight approximation is a result of the $K$-factor dependent
$r$. When the $K$-factor is 0 or 5 dB, it can be found that
$\delta\geq 0$ and the distribution of $\bigtriangleup_1$
$f_{\bigtriangleup_1}$ is asymmetric. Thus we let $r=1$. On the
other hand, when the $K$-factor is 10 or 15 dB,
$f_{\bigtriangleup_1}$ is unimodal and symmetric simultaneously.
In addition, $\delta\geq 2/\sqrt{3}$. Thus we set $r=2/9$ to
further tighten the original Cantelli's inequality. The slightly
over-designed $\alpha_1$ ensures that the primary user's outage
probability constraint is met at the cost of the CR user's rate.

\begin{figure}[htp]
\centering
\epsfig{file=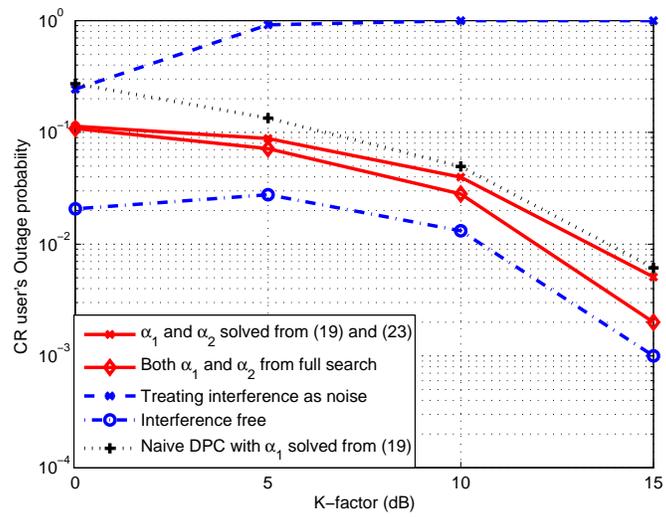,width=0.6\textwidth}
\caption{Comparison of the outage probabilities with $\alpha_2$
from full search and the approximation with $P_c=P_p=10$ and
$\sigma_{Z_s}^2=\sigma_{Z_p}^2=1$.}
\label{Fig_compare_outage_prob}
\end{figure}

The CR user's outage probabilities with $\alpha_1$ and $\alpha_2$
both obtained from approximations and full searches are given in
Fig. \ref{Fig_compare_outage_prob}. Recall that $\alpha_1$ is
slightly over-designed at medium and high $K$-factors. This
reduces the CR's power for its own signal, and increases the
outage probability. On the other hand, when the $K$-factor is
small, the worse channel condition dominates the performance. In
addition, we can find that the approximation from Alzer's bound
\cite{Alzer_incomplete_gamma_function} of the incomplete gamma
function is good for the considered $K$-factors. Naive DPC again
incurs a large rate loss in the low to medium $K$-factor region.
It can be seen that if the interference is treated as noise, the
outage probability is much worse than those of the proposed
precoding methods. Similar to the fast fading case, there is a gap
between the outage probabilities of the full and partial CSIT
cases. Since the optimal precoding strategy is still an open
problem, the minimum possible gap is unknown. The non-decreasing
outage probability curves of treating interference as noise and
full CSIT are due to the facts that the considered target rates
are not identical for different $K$-factors.

\vspace{-3mm}
\subsection{Lattice implementation with finite codelength }\label{sec_lattice_simu} 
\begin{figure}[htp]
\centering \epsfig{file=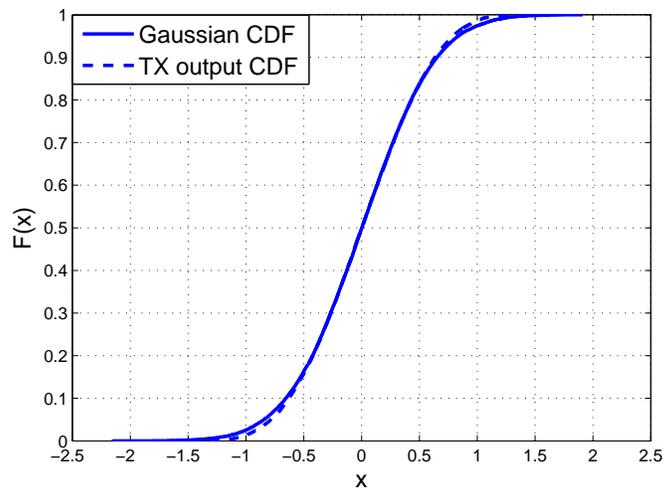,
width=0.6\textwidth} \caption{The Gaussianity of the CR
transmitted signal using the nested lattice code.}
\label{Fig_TX_distr}
\end{figure}

\vspace{-2mm}
We investigate the error performance of lattice precoding with a
reasonable codeword length (and decoding latency) at large enough
SNR. A pair of self-similar nested lattices is used, with the
Gosset lattice $E_8$ \cite{Conway_constructionA}, which has the
densest packing in 8-dimension, as the fine lattice $\Lambda_c$.
The reason to use the Gosset lattice $E_8$ is that in moderate
dimensions, it is well-known that the best lattices in terms of
coding gain are also good quantizers. That is, the lattice can
also be good $\Lambda_q$. The coarse lattice $\Lambda_q$ is
generated by $\Lambda_q=Q\Lambda_c$ where the coding rate is given
by $R=2\log Q$. Recall that transmitted signals is assumed to be
Gaussian in previous derivations.
The Gaussianity of the signals at the proposed lattice-based CR
transmitter output with a reasonably short codeword length $T=4$
is shown in Fig. \ref{Fig_TX_distr}. It is clear that the
transmitter output distribution is almost Gaussian.

\begin{figure}[htp]\centering
\epsfig{file=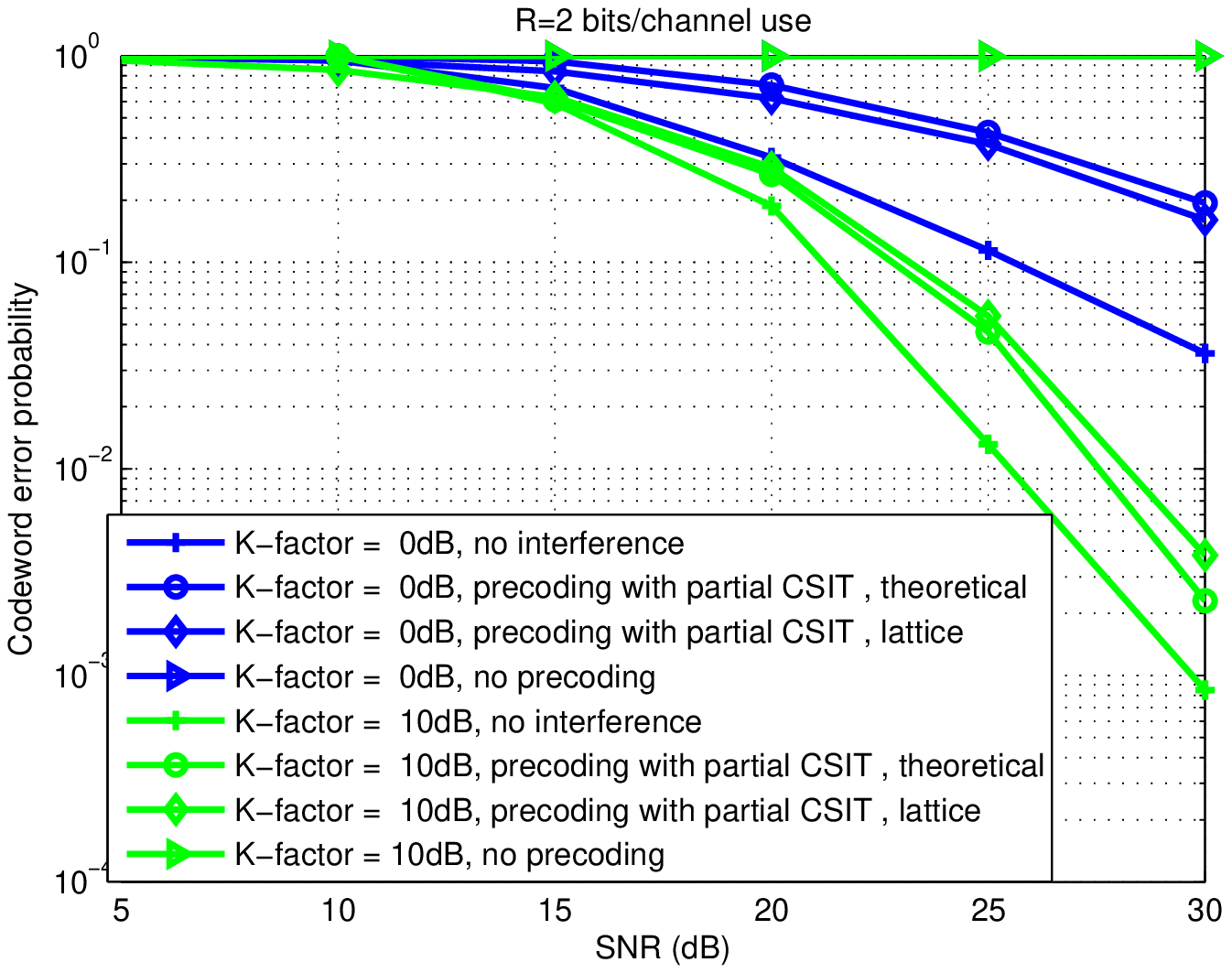,
width=0.6\textwidth} \caption{Comparison of the outage
probabilities with rate= 2 bpcu.} \label{Fig_PoutR2}
\end{figure}

\begin{figure}[htp]\centering
\epsfig{file=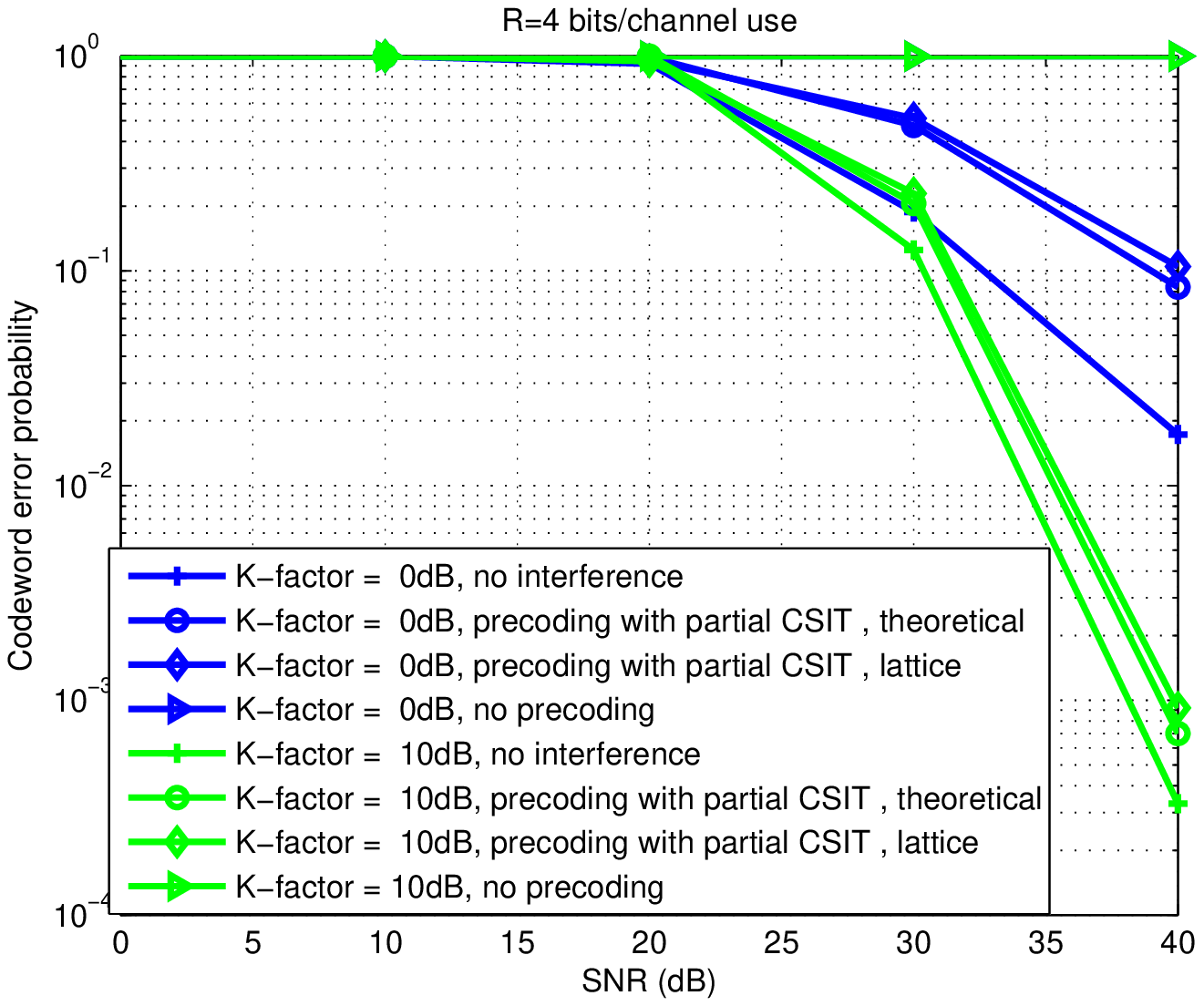,
width=0.6\textwidth} \caption{Comparison of the outage
probabilities with rate= 4 bpcu.} \label{Fig_PoutR4}
\end{figure}

In the simulation of codeword error probabilities we consider two
$K$-factors: 0dB and 10dB and two code rates: 2 and 4 bits per
channel use. The noise variance is normalized to 1 and the primary
user's power constraint is assumed to be $P_p=100$. The LA-GPC
curves are obtained from the lattice-code achievable rate with
filters specified in Theorem \ref{theorem_nested_DPC_LA}. A Fano
sequential-decoding based lattice decoder \cite{latticemurugan200}
is used to solve (\ref{Eq_2_minEuldecoder}). We compare the
lattice coding results with the theoretical results with random
Gaussian codebooks. The basis for the comparison is that the
codeword error probability is approximately the same as the outage
probability. The results are shown in Fig.~\ref{Fig_PoutR2} and
Fig.~\ref{Fig_PoutR4}. For comparison, we also show the results
without interference from the primary user and treating
interference as Gaussian noise. When the interference is treated
as noise, the codeword error rate is 1 regardless of the SNR and
$K$-factor ranges. It can be seen that in the considered
$K$-factor range, the performance of the proposed nested lattice
decoding is approximately the same as that derived theoretically.
This verifies the goodness of the used Gosset lattice code. We
also observe that in Fig. \ref{Fig_PoutR2} with $K$-factor = 0dB,
lattice precoding with partial CSIT performs slightly better than
the theoretical outage result. A similar phenomenon was also
reported in \cite{CaireLast}.

\vspace{-3mm}
\section{conclusion}\label{Sec_Conclusion}
In this paper we considered the cognitive radio channel with
partial CSIT. Using the linear-assignment Gel'fand-Pinsker coding,
we proposed semi-analytical methods for finding the relaying
ratios and the precoding coefficients for both fast and slow
fading channels. Asymptotic analysis showed that the relaying
ratios and the precoding coefficients obtained by the proposed
methods converge to those with full CSIT when the $K$-factor
approaches infinity. Simulation results showed that the proposed
semi-analytical parameter design methods perform almost as well as
exhaustive search. We also used nested-lattice coding and decoding
to realize the precoding in practice.  Simulation results showed
that the nested-lattice coding scheme can effectively reach the
achievable rate of the CR channel with partial CSIT at high SNR.

\appendix
\subsection{Proof of Theorem \ref{Th_1}}\label{App_1}
After applying the expectation operation to
(\ref{Eq_primary_matrix_form}) we have
\begin{align}
E[R(H_{11},H_{12})]&=E[\log(1+\underline{H}^H_p\mathbf{S}\underline{H}_p)]-E[\log(1+\underline{H}^H_p\mathbf{Q}\underline{H}_p)]\notag\\
&\triangleq
E[\log(1+\epsilon_1)]-E[\log(1+\epsilon_2)],\label{Eq_Appr_ergo_R}
\end{align}
where $\mathbf{S}\triangleq\mathbf{P}+\mathbf{Q}$,
$\varepsilon_1\triangleq
\underline{H}^H_p\mathbf{S}\underline{H}_p$, and
$\varepsilon_2\triangleq
\underline{H}^H_p\mathbf{Q}\underline{H}_p$. To guarantee that
(\ref{Eq_EG_Constriant}) is valid, we expand
$f(\varepsilon_1)\triangleq\log(1+\varepsilon_1)$ and
$g(\varepsilon_2)\triangleq\log(1+\varepsilon_2)$ by the Taylor
series with different orders, respectively. The reason is as the
following. We first expand $\log(1+\varepsilon_1)$ by the $k$-th
order Taylor series respect to the mean of $\varepsilon_1$,
$\mu_{\varepsilon_1}$, as
\begin{equation}\label{Eq_Taylor_expansion_of_R}
\log(1+\varepsilon_1)=\log(1+\mu_{\varepsilon_1})-\log
e\Sigma_{n=2}^k\frac{(-1)^n}{n}\frac{(\varepsilon_1-\mu_{\varepsilon_1})^n}{(1+\mu_{\varepsilon_1})^n}+o_1\triangleq
f_k(\varepsilon_1)+o_1,
\end{equation}
where $f_k(\varepsilon_1)$ is the truncated $k$-th order Taylor
expansion and $o_1$ is the Lagrange remainder. After applying the
expectation operator to (\ref{Eq_Taylor_expansion_of_R}) and
truncating the remainder $o$ we have
\begin{equation}E[f_k]=
\log(1+\mu_{\varepsilon_1})-\log
e\Sigma_{n=2}^k\frac{(-1)^n}{n}\frac{E[(\varepsilon_1-\mu_{\varepsilon_1})^n]}{(1+\mu_{\varepsilon_1})^n},
\end{equation}
where $E[f_k(\varepsilon_1)]$ is further simplified as $E[f_k]$.
Since the Taylor expansion of the logarithmic function $f$ is an
alternating series, and
$E[(\varepsilon_1-\mu_{\varepsilon_1})^{n+1}]>0,\;\forall
n\in\mathds{N}$, from \cite{Mathai}, we have the following
relations: $E[f_1]>E[f_3]>E[f_5]>\cdots>E[f]$ and
$E[f_2]<E[f_4]<E[f_6]<\cdots<E[f]$. Similar properties can be
found for $g(\varepsilon_2)$. The truncated first order Taylor
expansion of $E[\log(1+\varepsilon_2)]$ can be derived in the same
way. From the above we know that
$E[f]-E[g]>E[f_{2i}]-E[g_{2j-1}]$. Thus solving the following
equality
\[
E[f_{2i}]-E[g_{2j-1}]=R_{ergodic}^P
\]
makes the constraint (\ref{Eq_EG_Constriant}) valid where
$i,j\in\mathds{N}$. Here we choose $i=j=1$ for simplicity. After
some manipulations we have (\ref{Eq_appr_ergo_al}), where
$\mu_{\varepsilon_1},\,\sigma^2_{\varepsilon_1},$ and
$\mu_{\varepsilon_2}$, the mean, variances of $\varepsilon_1$, and
the mean of $\varepsilon_2$, can be found according to
\cite{Mathai}.

\subsection{Proof of Proposition \ref{Th_2}}\label{App_2}
With the fact that $c_0(\mathbf{P}+\mathbf{Q})-\mathbf{D}$ is
positive semi-definite and $c>0$, $B(\alpha_2)$ is positive. Then
we can generalize the result in
\cite{Lieberman_Laplace_approximation}, which is aimed for the
ratio of quadratic forms, to the ratio of general quadratic
forms\footnote{"General" means that there is not only the term of
the quadratic form, but also terms of  lower orders.} in
(\ref{Eq_general_moment_of_ratio_of_quadratic_form}). Therefore,
we have
\[
E\left[\frac{B'(\alpha_2)}{B(\alpha_2)}\right]\simeq
\frac{E[B'(\alpha_2)]}{E[B(\alpha_2)]}=0.
\]
Since $B(\alpha_2)$ is a second order polynomial of $\alpha_2$, to
proceed, we only need to consider the numerator, that is,
\[
E[B'(\alpha_2)]=0.
\]
With the fact that $\sigma^2_{22}+|\mu_{22}|^2=1$, the resulting
precoding coefficient with the statistics of CSIT is as
(\ref{Eq_alpha2_1}).

\subsection{Proof of Theorem \ref{Th_3}}\label{App_3}
From (\ref{Eq_primary_matrix_form}), (\ref{Eq_PR_rate_constraint})
and (\ref{Eq_delta1}), the optimal $\alpha_2$ will make
\begin{equation}\label{Eq_outage_constraint}
P(\bigtriangleup_1 \geq\,\frac{1}{2^{R^P}-1})= P_{out}^{P}.
\end{equation}
To make this problem analytically tractable, we relax the equality
constraint by inequality to find the sub-optimal solution. With
the non-negativity of the transmission rate, the Cantelli's
inequality \cite{Cantelli_inequality} is a good candidate for
finding the solution. Considering additional properties of the
distribution $f_{\bigtriangleup_1}$, such as the modality and
symmetry, the Cantelli's inequality can be further tightened
\cite{modified_Cantelli}. Thus we propose the use of the following
modified Cantelli's inequality to relax
(\ref{Eq_outage_constraint}) as
\begin{equation}
P(\bigtriangleup_1\geq
\mu_{\bigtriangleup_1}+\delta\sigma_{\bigtriangleup_1})\leq
\frac{r}{1+\delta^2}.\label{Eq_Cantelli_general}
\end{equation}
 According to different $f_{\bigtriangleup_1}$
resulting from different channel conditions, $r$ in
(\ref{Eq_Cantelli_general}) can be categorized as
\cite{modified_Cantelli}
\begin{equation} \label{eq_r}
r=\left\{\begin{array}{l}1,\,\mbox{ if }\delta\geq 0,\\
\frac{2}{9},\,\mbox{ if }\delta\geq\frac{2}{\sqrt{3}}\; \mbox{and}
\; f_{\bigtriangleup_1}\mbox{ is unimodal and
symmetric.}\end{array}\right.
\end{equation}
Comparing the left hand side of the inequalities in
(\ref{Eq_outage_constraint}) and (\ref{Eq_Cantelli_general})
results in
\begin{equation}\label{Eq_find_delta}
\mu_{\bigtriangleup_1}+\delta\sigma_{\bigtriangleup_1}=\frac{1}{2^{R^P}-1}.
\end{equation}
Also by comparing the right hand side of the inequalities in
(\ref{Eq_outage_constraint}) and (\ref{Eq_Cantelli_general}) we
have $\delta=\sqrt{r/P_{out}^P-1}$. Substituting $\delta$ into
(\ref{Eq_find_delta}), we have (\ref{Eq_cantelli_approximation}).
To determine $r$, we can simply substitute $r=1$ or $r=2/9$ to see
whether the resulting $\delta$ meets the constraints (\ref{eq_r}).
The constraints of modality and (or) symmetry which may be
satisfied under different channel conditions are discussed in
Section \ref{Sec_simulation}. Thus we can solve the outage
probability constraint by finding $\alpha_1$ which satisfies the
equality in (\ref{Eq_cantelli_approximation}). We now discuss the
validity of (\ref{Eq_PR_rate_constraint}). As for the effect of
the modified Cantelli's inequality, note that the modified
Cantelli's upper bound is larger than the real probability value
\cite{modified_Cantelli}. Since we equate the upper bound to the
target outage probability to solve $\alpha_1$, the resulting
$\alpha_1$ will be larger than the optimal one from
(\ref{Eq_outage_constraint}).

\subsection{Proof of Theorem \ref{Thereom_slow_fading alpha_2}}\label{App_4}
First, substituting (\ref{Eq_CR_rate_matrix}) into
(\ref{Eq_CR_Pout_constraint}), and after some manipulations we
have the following optimization problem
\begin{equation} \label{Eq_alpha_2_slow}
\min_{\alpha_2}\,P(z\triangleq\underline{H}_c^{\mathrm{H}}\mathbf{E}\underline{H}_c=\underline{H}_2^{\mathrm{H}}\mathbf{E}_2\underline{H}_2
<cd-1),
\end{equation} where $\mathbf{E}_2=\Sigma^{1/2}\mathbf{E}\Sigma^{1/2}$, and
$\underline{H}_2\sim\mathpzc{CN}(\Sigma^{-1/2}\mu,\mathbf{I}_2)$.
The transformation from $\mathbf{E}$ to $\mathbf{E}_2$ is to make
$\underline{H}_2$ have an identity covariance matrix for the
convenience of the following analysis. After substituting the
eigen decomposition
$\mathbf{E}_2=\mathbf{v}^{\mathrm{H}}\Lambda\mathbf{v}$ into $z$,
we have
\[
z= \underline{H}_3^{\mathrm{H}}\Lambda\underline{H}_3 \sim
\lambda_1\chi^2(2,2|\mu_{31}|^2)+\lambda_2\chi^2(2,2|\mu_{32}|^2),
\]
where $\underline{H}_3\sim \mathpzc{CN}(\mu_3,\mathbf{I}_2)$, $
\mu_3 = [\mu_{31}$, $\mu_{32}]^{T} \triangleq
\mathbf{v}\Sigma^{-1/2}\mu$, and $\lambda_1$ and $\lambda_2$ are
the eigenvalues of $\mathbf{E}_2$. The notation $\chi^2 (p,q)$
denotes the noncentral chi-square distribution with degree of
freedom $p$ and non-centrality $q$. Note that the objective
function (\ref{Eq_alpha_2_slow}) is a function of $\alpha_2$. That
is, $\lambda_1,\,\lambda_2,\,\mu_{31},$ and $\mu_{32}$ are
controlled by $\alpha_2$.

To calculate the outage probability (\ref{Eq_alpha_2_slow}), we
resort to approximating $z$ by a scaled single central chi-square
random variable \cite{Kendall_Chi_appr} as
\begin{equation}\label{Eq_Chi_square_approximation}
z \sim
\lambda_1\chi^2(2,2|\mu_{31}|^2)+\lambda_2\chi^2(2,2|\mu_{32}|^2)\simeq
v\chi^2(w).
\end{equation}
Recall that the chi-square distribution is a special case of the
gamma distribution. Thus $w$ does not have to be an integer. The
parameters $v$ and $w$ should be chosen such that both sides of
(\ref{Eq_Chi_square_approximation}) have the same first and second
moments \cite{Kendall_Chi_appr} as
\begin{align}
v w&=\lambda_1(1+|\mu_{31}|^2)+\lambda_2(1+|\mu_{32}|^2)\label{Eq_vw},\\
2v^2w&=\lambda_1^2(1+2|\mu_{31}|^2)+\lambda_2^2(1+2|\mu_{32}|^2).\notag
\end{align}
With the fact
\begin{align}
\lambda_1+\lambda_2&=\mbox{tr}(\mathbf{E}_2)=\mbox{tr}(\Sigma\mathbf{E}),\label{Eq_vw_1}\\
\lambda_1|\mu_{31}|^2+\lambda_2|\mu_{32}|^2&=\mu_3^{\mathrm{H}}\Lambda\mu_3=\mu^{\mathrm{H}}\Sigma^{\frac{-1}{2}}\mathbf{v}^{\mathrm{H}}\Lambda\mathbf{v}\Sigma^{\frac{-1}{2}}\mu=\mu^{\mathrm{H}}\mathbf{E}\mu,\label{Eq_vw_2}
\end{align}
we can find the value of $vw$. Similarly, from
\begin{align}
\lambda_1^2+\lambda_2^2&=\mbox{tr}(\mathbf{E}_2\mathbf{E}_2)=\mbox{tr}(\Sigma\mathbf{E}\Sigma\mathbf{E})\\\notag,
\lambda_1^2|\mu_{31}|^2+\lambda_2^2|\mu_{32}|^2&=\mu^{\mathrm{H}}\Sigma^{\frac{-1}{2}}\mathbf{E}_2\mathbf{E}_2\Sigma^{\frac{-1}{2}}\mu=\mu^{\mathrm{H}}\mathbf{E}\Sigma\mathbf{E}\mu.
\end{align}
we can find the value of $2v^2w$. Then $v$ and $w$ can be
expressed as (\ref{Eq_vw_outage}). After that, solving
(\ref{Eq_alpha_2_slow}) with (\ref{Eq_Chi_square_approximation})
is equivalent to solving
\[
\min_{\alpha_2}\,\,\frac{\int_0^{\frac{cd-1}{2v(\alpha_2)}}x^{\frac{w(\alpha_2)}{2}-1}e^{-t}dt}{\Gamma(\frac{w(\alpha_2)}{2})},
\]
which can be further represented as (\ref{Eq_alpha2_bound}).
\subsection{Proof of Corollary \ref{Corollary_1}}\label{App_corollary1}
From \cite{Alzer_incomplete_gamma_function} we know that the
incomplete gamma function can be bounded as following
\[
(1-e^{-sx})^a<\frac{\gamma(a,x)}{\Gamma(a)},
\]
where
\begin{equation}
s=\left\{\begin{array}{ll} 1,\,&0<a<1\\
(\Gamma(1+a))^{\frac{-1}{a}},\,&a>1.
\end{array}
\right.
\end{equation}
Let $a=\frac{w(\alpha_2)}{2}$ and $x=\frac{cd-1}{2v(\alpha_2)}$,
we then have (\ref{Eq_alpha22}).

\subsection{Proof of Corollary \ref{L_1}}\label{App_5}
As the $K$-factor$\rightarrow \infty$, the distribution of $z$
defined in (\ref{Eq_alpha_2_slow}) becomes a delta function with a
nonzero value at its mean $\mu_z$. In the following we need to
find $\alpha_2$ maximizing the mean value of $z$ which can be
approximated by $vw$, as shown in
(\ref{Eq_Chi_square_approximation}). Since $c$ in
(\ref{Eq_alpha_2_slow}) is a function of $\alpha_2$, we must take
it into consideration. The problem becomes to maximize $\mu_z-cd$.
As both $\mu_{21}$ and $\mu_{22}\rightarrow 1$, and
$\sigma_{21}^2$ and $\sigma_{22}^2\rightarrow 0$ at very high
$K$-factor, from (\ref{Eq_vw}), (\ref{Eq_vw_1}), and
(\ref{Eq_vw_2}), we have
\begin{align}
\lim_{K\rightarrow}\mu_z&=
vw=\mu^{\mathrm{H}}\mathbf{E}\mu+\mbox{tr}(\Sigma\mathbf{E})\simeq
e_{11}+e_{12}+e_{21}+e_{22},\notag
\end{align}
where $e_{11},\,e_{12},\,e_{21},$ and $e_{22}$ are the four
entries of $\mathbf{E}$. Since $\mu_z-cd$ is a quadratic formula
of $\alpha_2$, we can take derivative to find the maximum as
following
\begin{align}
\frac{1}{d}\frac{\partial }{\partial
\alpha_2^*}(\mu_z-cd)&=-\left((1-\alpha_1)P_c+1\right)P_p\alpha_2+(P_p+\sqrt{\alpha_1P_cP_p})(1-\alpha_1)P_c=0.\notag
\end{align}
Then $\alpha_2$ can be found as
\begin{equation}
\lim_{K\rightarrow\infty}\alpha_2=\frac{(1+\sqrt{\frac{\alpha_1
P_c}{P_p}})(1-\alpha_1)P_c}
{(1-\alpha_1)P_c+1}=(1+\sqrt{\frac{\alpha_1
P_c}{P_p}})\alpha_2^{MMSE},
\end{equation}
which is the same as that of the non-fading case.

\bibliographystyle{IEEEtran}
\bibliography{IEEEabrv,codeSN,DPC_PAPR2}
\newpage


\end{document}